\documentstyle[prc,preprint,float,aps,epsfig,amsmath,rotating]{revtex}

\begin{document}

\draft

\title{Shell model study of $^{40}$Ca muon capture
and the $(0^+ , 0)$$\rightarrow$$(0^- , 2626)$ 
axial charge transition}

\author{T.P.~Gorringe}

\address{Department of Physics and Astronomy,
University of Kentucky, Lexington, KY 40506}

\date{\today}
\maketitle

\begin{abstract}

We report results from shell model studies of
muon capture on $^{40}$Ca to low-lying levels
of $^{40}$K. We discuss the comparison between 
calculated capture rates, measured capture rates 
and analogous transitions in ($e$,$e^{\prime}$) scattering 
in terms of the particle-hole
structure of the $^{40}$Ca-$^{40}$K nuclei.
We highlight the
$^{40}$Ca$(0^+ , 0)$$\rightarrow$$^{40}$K$(0^- , 2626)$ 
axial charge transition and its sensitivity to 
the induced pseudoscalar coupling $g_p$
of the proton's weak interaction.
In addition, we address the hindrance of 
unique first-forbidden transitions
due to particle-hole interactions
and the emergence of
allowed Gamow-Teller transitions 
due to ground state correlations.
Lastly, we examine the longitudinal alignment
of $^{40}$K recoils following muon capture,
and discuss this possibility for independently 
determining the induced coupling $g_p$.
\end{abstract}


\section{Introduction}
\label{s: introduction}

Studies of muon capture on complex nuclei
have spanned nearly fifty years.
The early work \cite{Ma61,Cu61,Te62}
was devoted to establishing
the universal V-A character  
of nuclear muon capture,
while later work 
has focused on
induced currents \cite{Go04}, 
second-class currents \cite{Ho76,Le77}
and non $V$-$A$ interactions \cite{Sh02}. 
Muon capture
is also a valuable window on
particle-hole excitations and
spin-isospin modes
in nuclei \cite{Me01},
it complementing the information 
that is obtained from beta-decay,
electron scattering
and charge exchange reactions.

Of particular interest in nuclear muon capture 
is the induced pseudoscalar coupling $g_p$
of the proton's weak interaction.
For the free proton, the coupling $g_p$
is predicted to few-percent accuracy by 
symmetry arguments and thereby represents 
an important test of low energy QCD \cite{Go58,Be94,Fe97}.
For the bound proton,
its medium modification 
is sensitive to effects that range from pion exchange currents
and $\Delta$-hole excitations to
partial restoration of chiral symmetry 
\cite{De76,De94,Rh84}.
Indeed, some suggestions of a large A-dependent renormalization
of the induced coupling $g_p$ have been 
published in the literature \cite{De76,Gm88}.

Measday and Stocki \cite{Me06} have recently published 
new experimental results on $\gamma$-ray spectra 
from muon capture on $^{40}$Ca
that include determinations of
partial rates and rate limits
for numerous $^{40}$Ca($\mu$,$\nu$)$^{40}$K transitions.
Their data are noteworthy
as the heaviest nucleus on which ($\mu$,$\nu$) reactions
have been clearly identified.
Their data show evidence of excitations that range
from allowed Gamow-Teller transitions
to highly $\ell$-forbidden transitions.

Most important
is their observation of the 
$^{40}$Ca$(0^+ , 0)$ $\rightarrow$ $^{40}$K$(0^- , 2626)$ axial charge
transition. In nuclear beta decay such $0^+$$\leftrightarrow$$0^-$ transitions
have been extensively studied --
both experimentally and theoretically --
and nowadays represent our clearest evidence 
for exchange currents in complex nuclei. 
In muon capture the only known example 
of a $0^+$$\leftrightarrow$$0^-$ first forbidden transitions 
was the celebrated $^{16}$O$(0^+, 0)$$\rightarrow$$^{16}$N$(0^-, 120)$ 
transition, it having been studied by many authors
in the context of the coupling $g_p$.
In light of suggestions of an A-dependent renormalization
of the induced coupling $g_p$, 
the $^{40}$Ca$(0^+ , 0)$ $\rightarrow$ $^{40}$K$(0^- , 2626)$
transition represents a valuable data point
for experimentally accessing $g_p$ in medium-weight nuclei.

Also interesting are two unique first forbidden transitions
and one allowed Gamow Teller transition:
$^{40}$Ca$(0^+ , 0)$ $\rightarrow$ $^{40}$K$(2^- , 800 )$,
$^{40}$Ca$(0^+ , 0)$ $\rightarrow$ $^{40}$K$(2^- , 2047 )$
and $^{40}$Ca$(0^+ , 0)$ $\rightarrow$ $^{40}$K$(1^+ , 2290 )$.
One motivation for investigating unique first-forbidden transitions 
stems from the large quenching of such transitions
in nuclear $\beta$-decay,
such quenching arising 
from the coherent effects of
the repulsive $T$=$1$ particle-hole interaction.
One motivation for investigating allowed Gamow-Teller transitions
stems from their absence in a simple closed-shell description
of doubly-magic $^{40}$Ca,
their emergence arising 
from the multi-particle, multi-hole admixtures
in the $^{40}$Ca ground state.
In addition, the $^{40}$Ca$(0^+ , 0)$ $\rightarrow$ $^{40}$K$(2^- , 800 )$
and $^{40}$Ca$(0^+ , 0)$ $\rightarrow$ $^{40}$K$(2^- , 2290 )$
transitions have well known analogs
in ($e$,$e'$) inelastic scattering \cite{Bu82,Og84,Pe95}
and charge exchange reactions \cite{Ta84,Ch86}.
The comparison between $( \mu^- , \nu )$ capture rates
and analogous ($e$,$e^{\prime}$) transitions thus provides
a valuable cross-check
for the  $^{40}$Ca($\mu$,$\nu$)$^{40}$K data
and a valuable comparison
for the $^{40}$Ca($\mu$,$\nu$)$^{40}$K calculations.

The $^{40}$Ca($\mu$,$\nu$)$^{40}$K data 
are also interesting in the broader context 
of the shell structure of the $A \sim 40$ region.
Warburton and co-workers have published
a series of articles \cite{Wa86,Wa88a,Wa88b,Wa88c,Wa90} 
on mass $34 \leq A \leq 50$ nuclei
that include comprehensive studies 
of $\Delta$$J^{\pi} = 0^-$ $\beta$ decays 
and their enhancement due to exchange currents effects,
$\Delta$$J^{\pi} = 2^-$ $\beta$ decays
and their hindrance due to core polarization effects,
and the evolving nuclear structure 
of the $A \sim 40$ mass region.
Weak transition rates on doubly magic $^{40}$Ca
are the archetypal benchmark
for such model calculations.

Herein we report results from
shell model studies of muon capture on $^{40}$Ca. 
In Secs.\ \ref{s: shell calculations} and \ref{s: capture calculations} respectively 
we describe the details of the shell model
calculations and the capture rate calculations.
In Sec.\ \ref{s: rates} we compare the 
calculated ($\mu$,$\nu$) transition rates
with the recent data of
Measday and Stocki \cite{Me06} and the earlier data 
of Igo-Kemenes {\it et al.}\ \cite{Ig71} in the context 
of the particle-hole structure 
of the $^{40}$Ca-$^{40}$K nuclear wavefunctions.
In particular, we discuss
the emergence of the ($0^+$,0)$\rightarrow$(1$^+$,2290) 
allowed Gamow-Teller transition, 
the hindrance of the $0^+$$\rightarrow$2$^-$ unique first-forbidden
transitions, and the  sensitivity to the induced pseudoscalar coupling $g_p$ 
of the $^{40}$Ca$(0^+ , 0)$$\rightarrow$$^{40}$K$(0^- , 2626)$ 
axial charge transition.
Finally, in Sec.\ \ref{s: alignments}
we briefly consider the longitudinal alignment
of $^{40}$K recoils following $^{40}$Ca capture,
and the dependence
of the alignment
on the coupling $g_p$.

\section{Shell model calculations}
\label{s: shell calculations}

The simplest desription of muon capture from the
$^{40}$Ca ground state to the low-lying $^{40}$K 
states involves a fully filled sd-shell initial state
and various (sd)$^{-1}$(pf)$^{1}$ particle-hole
final states. This model infers 
a significant number of forbidden transitions but
no allowed transitions as all spin-orbit partners
in the $^{40}$Ca ground state are fully filled.
It implies that data for allowed
transitions will have interesting sensitivity to 
the multi-particle, multi-hole admixtures
in the $^{40}$Ca g.s.\ wavefunction
and data for forbidden transitions will have 
interesting sensitivity to the particle-hole interaction
in the $A \sim 40$ mass region.

\begin{figure}
\begin{center} 
\mbox{\epsfig{figure=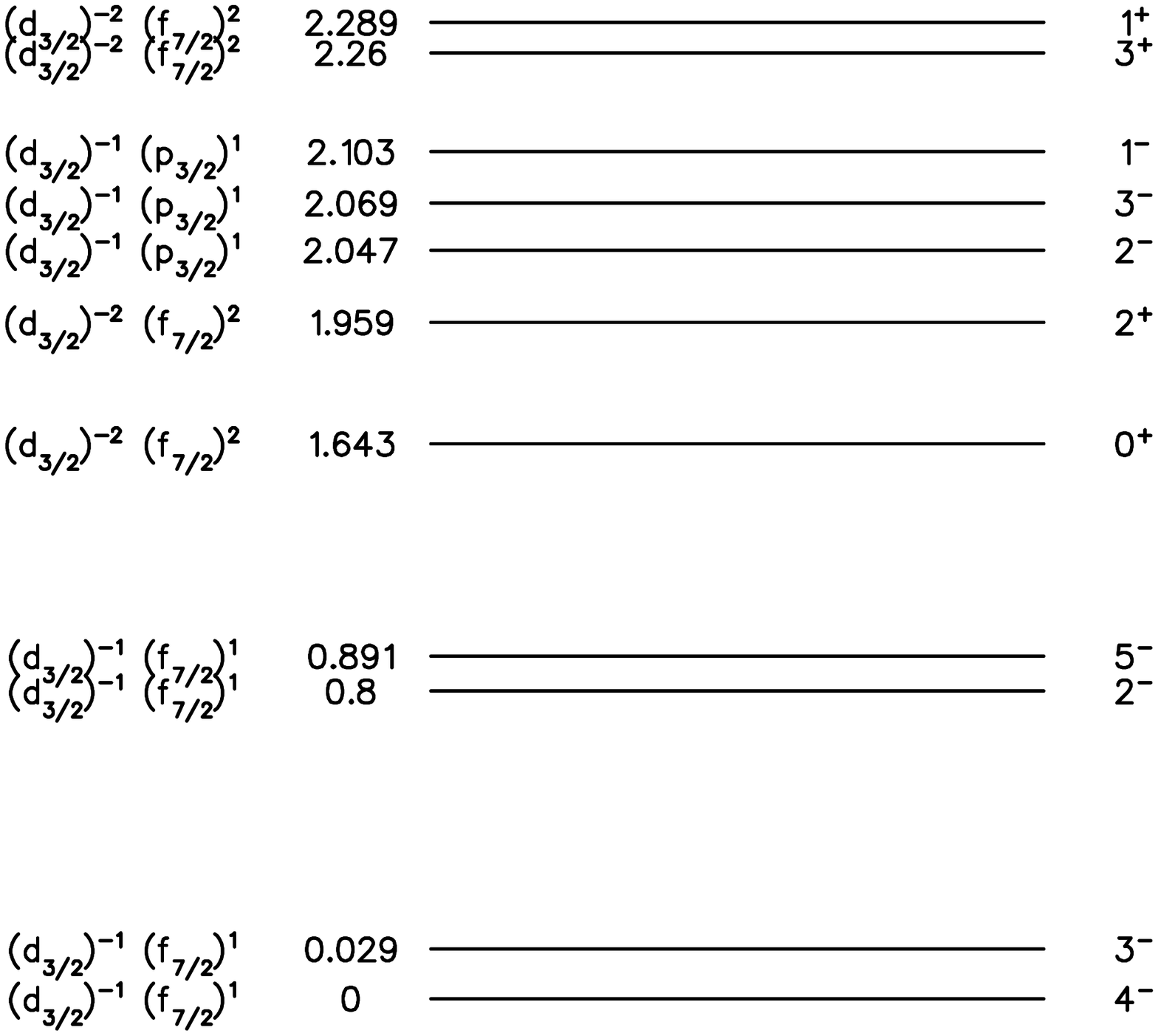,height=16.0cm,angle=90}}
\end{center}
\caption{Summary of assignments of model states to
experimental states for the low-lying $^{40}$K levels.
Shown are the spin-parities, excitation
energies ,and dominant particle-hole configurations
for the low-lying $^{40}$K states.}
\label{f: elevel}
\end{figure}

The $^{40}$K level diagram of Fig.\ \ref{f: elevel}
reveals one negative
parity quadruplet with spin-parities $J^{\pi} = 4^-, 3^-, 2^-$ and $5^-$
and excitation energies $0, 30, 800$ and  $891$~keV
and another negative parity quadruplet 
with spin-parities $J^{\pi} =  2^-, 3^-, 1^-$ and $0^-$
and excitation energies $2047, 2070, 2104$ and $2626$~keV.
The quadruplets are consistent with  
low-lying negative parity $^{40}$K states 
involving $(d_{3/2})^{-1}(f_{7/2})^{1}$
particle-hole configurations, which yield the 
$J^{\pi} = 2^-$-$5^-$ quadruplet,
and $(d_{3/2})^{-1}(p_{3/2})^{1}$
particle-hole configurations, which yield the
$J^{\pi} = 0^-$-$3^-$ quadruplet.
In addition, the four lowest-lying positive parity states,
with spin-parities $J^{\pi} =  0^+, 2^+, 3^+$ and $1^+$  and excitation
energies $1632, 1959, 2260$ and $2290$~keV, were identified
by Davis {\it et al}.\ \cite{Da73} 
as consistent with  expectations of
low-lying $(d_{3/2})^{-2}(f_{7/2})^{2}$ excitations 
with the two  $f_{7/2}$-particles coupled to $J = 0$ 
and the two  $d_{3/2}$-holes coupled to $J = 0$-$3$.
At higher excitation energies the increasing level density 
makes meaningful assignments
between model configurations and experimental levels
much more difficult.

Below we describe
the model spaces and residual interactions we
have employed in our calculations
of the muon capture rates
and the $^{40}$K recoil alignments.
Our basic philosophy 
was to utilize well-known, well-tested models 
``off-the-shelf'', {\it i.e.}\
not fine tuning 
any model parameters.
Our model calculations
were performed utilizing
the OXBASH code \cite{Ox86}.

\subsection{$A = 40$ model spaces and residual interactions}

Our primary calculations of 
$^{40}$Ca$( \mu, \nu )$$^{40}$K capture
were conducted using the 
sd-pf model space and the WBMB interaction \cite{Wa89}.
The WBMB interaction comprises 
the sd interaction of Wildenthal \cite{Wi84}, 
the pf interaction of McGory \cite{Mc73}, 
and a modified Millener-Kurath interaction for the
cross-shell matrix elements \cite{Mi75}.
As discussed in Ref.\ \cite{Wa86},
the sd-pf shell energy gap
was empirically adjusted to reproduce
the $A = 35$-$41$ binding energies
and the $d_{3/2}$$f_{7/2}$$\leftrightarrow$$d_{3/2}$$f_{7/2}$,
$d_{3/2}$$p_{3/2}$$\leftrightarrow$$d_{3/2}$$p_{3/2}$ 
two-body matrix elements
were empirically adjusted to reproduce the 
low-lying $A = 40$ particle-hole states.

Two versions of WBMB calculations were carried out.
In both versions of the WBMB model
the negative-parity $^{40}$K states comprised
one particle in the fp shell and one hole in the sd shell,
{\it i.e.}\ $(sd)^{23}$$(pf)^1$ configurations,
and the positive-parity $^{40}$K states comprised
two particles in the fp shell and two holes in the sd shell,
{\it i.e.}\ $(sd)^{22}$$(pf)^2$ configurations.
However, to enable the investigation of ground state correlations
we employed $^{40}$Ca wavefunctions 
with both a pure $(sd)^{24}$ configuration,
our so-called $0$~$\hbar$$\omega$ WBMB calculation,
and a mixed $(sd)^{24}$+$(sd)^{22}$$(pf)^2$ configuration, 
our so-called ($0$+$2$)~$\hbar$$\omega$ WBMB calculation.
We caution the reader the WBMB interaction
was originally developed and generally applied
in pure N~$\hbar$$\omega$ model spaces, although
the effects of (0+2)~$\hbar$$\omega$ mixing
on hypothetical $^{40}$K$\rightarrow$$^{40}$Ca
beta-decays was discussed in Ref.\ \cite{Wa88a}.\footnote{The
correct treatment of $(0 + 2)$$\hbar$$\omega$ mixing
is a well-known difficulty in shell model calculations
See Ref.\ \cite{Wa86} for a discussion of $(0 + 2)$$\hbar$$\omega$ mixing 
in the $A \sim 40$ mass region.}

We also performed
calculations that employed a smaller $d_{3/2}$-$f_{7/2}$ model space
but allowed a richer 0-8~$\hbar$$\omega$ particle-hole spectrum.
These calculations used
the single particle energies and two-body matrix elements
of  Sakakura, Arima and Sebe \cite{Sa76} which is hereafter denoted 
as the SAS interaction.
This interaction was used in Ref.\ \cite{Sa76}
to study the effects of multi-particle, multi-hole
admixtures in $0^+$ levels of $^{40}$Ca.
While the smaller $d_{3/2}$-$f_{7/2}$ space
is obviously unable to fully describe
all low-lying $^{40}$K particle-hole levels
-- it missing the $f_{7/2}^{-1}$-$p_{3/2}^1$ states --
it renders some insight into the 
multi-particle, multi-hole components 
in the $^{40}$Ca ground state wavefunction
and their influence
on the $( \mu , \nu )$ transitions.

\subsection{$^{40}$K excitation energies and wavefunction configurations}

The WBMB calculations exactly reproduces
the excitation energies of the $^{40}$K lower-lying
$J^{\pi} = 4^-, 3^-, 2^-, 5^-$ quadruplet
and the $^{40}$K higher-lying $J^{\pi} = 2^-, 3^-, 1^-, 0^-$  quadruplet
due to the aforementioned adjustments
of  the $d_{3/2}$$f_{7/2}$$\leftrightarrow$$d_{3/2}$$f_{7/2}$,
$d_{3/2}$$p_{3/2}$$\leftrightarrow$$d_{3/2}$$p_{3/2}$
two-body matrix elements.
Within the WBMB model the 
$J^{\pi} = 4^-, 3^-, 2^-, 5^-$ quadruplet 
has $d_{3/2}^{-1}$-$f_{7/2}^1$ occupancies
that range from 92 to 98\% and
the $J^{\pi} = 2^-, 3^-, 1^-, 0^-$ quadruplet 
has $d_{3/2}^{-1}$-$p_{3/2}^1$ occupancies
that range from 75 to 99\% 
(the occupancies are consistent
with the compilation
of spectroscopic factors 
in Endt {\it et al}.\ \cite{En77}).
By comparison the WBMB model 
yields low-lying $^{40}$K positive parity states 
that are typically 1~MeV under-bound
and have somewhat exaggerated energy spacings --
although the model does reproduce the 
$J^{\pi} = 0^+, 2^+, 3^+, 1^+$ level ordering.
While the $J^{\pi} = 0^+, 2^+, 3^+$ states
have dominant $d_{3/2}^{-2}$$f_{7/2}^{2}$ character
the $J^{\pi} = 1^+$ state
is a more fragmented mixture 
of the permissible 2p-2h configurations.

The SAS model reproduces the excitation
energies of the lower-lying negative-parity $^{40}$K quadruplet
with spin-parities $J^{\pi} = 4^-, 3^-, 2^-, 5^-$
to accuracies of roughly 200~keV.
The corresponding SAS wavefunctions have their largest
contributions  
from $d_{3/2}^{-1}$$f_{7/2}^1$ configurations (43-55\%)
but with substantial contributions 
from $d_{3/2}^{-3}$$f_{7/2}^3$ configurations.(31-40\%).
The $J^{\pi} = 0^+, 2^+, 3^+, 1^+$ level ordering
is reproduced by the SAS model
although the $J^{\pi} =0^+, 2^+$ states
are over-bound by roughly 1~MeV.
All four positive parity states
have their largest contributions from 
$d_{3/2}^{-2}$$f_{7/2}^2$ configurations (42-61\%)
but with substantial contributions from
$d_{3/2}^{-4}$$f_{7/2}^4$ configurations (34-40\%).
Of course the SAS model and $d_{3/2}$-$f_{7/2}$ space
is unable to account for the higher-lying 
negative-parity $^{40}$K quadruplet
with its dominant $d_{3/2}^{-1}$-$p_{3/2}^1$ character.

\subsection{$^{40}$Ca ground state wavefunction and np-nh admixtures}
\label{s: 40Ca ground state}

The various calculations are quite different
in their descriptions
of the $^{40}$Ca g.s.
While the 0~$\hbar$$\omega$ WBMB calculation
assumes a simple closed-shell $^{40}$Ca g.s.\
the closed-shell occupancies 
are 72\% in the (0+2)~$\hbar$$\omega$ WBMB calculation 
and 46\% in the SAS calculation.
In addition, 
the (0+2)~$\hbar$$\omega$ WBMB calculation
and SAS calculation
have qualitatively different
np-nh wavefunction admixtures.
For example,
in the SAS calculation
the leading np-nh component
is the $d_{3/2}^{-2}$-$f_{7/2}^{2}$ configuration
with a 35\% occupancy
whereas in the (0+2)~$\hbar$$\omega$ WBMB calculation
the large np-nh fragmentation  
yields a $d_{3/2}^{-2}$-$f_{7/2}^{2}$ configuration
with only 2\% occupancy.

The different descriptions 
of the $^{40}$Ca g.s.\ 
-- in particular the differences
in closed-shell occupancies and
multi-particle, multi-hole admixtures --
is the leading cause
of the model-to-model variation
in the calculated capture rates. 
It is worthwhile noting the
negative-parity transitions and positive-parity transitions
are endowed with qualitatively different sensitivities
to the np-nh admixtures
in the $^{40}$Ca ground state.
This difference originates 
as negative-parity transitions 
are dominantly one-body transitions 
connecting the $^{40}$Ca closed-shell wavefunction component
and the $^{40}$K 1p-1h wavefunctions
while the positive-parity transitions
are dominantly one-body transitions 
connecting the $^{40}$Ca 2p-2h wavefunction components 
and the $^{40}$K 2p-2h wavefunction.
Consequently, the negative-parity transitions
are mostly sensitive to the magnitude
of the $^{40}$Ca 2p-2h admixtures
while the positive-parity transitions
are additionally sensitive to the configurations
in the $^{40}$Ca 2p--2h admixtures.

\section{Muon capture calculations}
\label{s: capture calculations}

In calculating the muon capture observables
and nuclear matrix elements
we used the formalism and notation
of Walecka \cite{Wa75} and Donnelly and Haxton \cite{Do79}
(for details see Ref.\ \cite{Jo96}).
Note that in $^{40}$Ca capture
each $0^+$$\rightarrow$$J_f^{\pi}$ transition
involves a single spin-parity multipole 
and all observables are governed 
by the nuclear matrix elements of two electroweak operators,
{\it i.e.} $\!{\hat{\cal L}}_J\!-\!{\hat{\cal M}}_J$
and ${\hat{\cal T}}^{el}_J\!-\!{\hat{\cal T}}^{mag}_J$.
In principle these nuclear matrix elements 
can involve the contributions of one-body, two-body and
many-body weak nucleonic currents.
Herein we assume these operators
can be represented by the sum of the
A one--body weak nucleonic currents,
{\it i.e.}\ the impulse approximation.
In such circumstances 
the nuclear matrix elements of
$\!{\hat{\cal L}}_J\!-\!{\hat{\cal M}}_J$
and ${\hat{\cal T}}^{el}_J\!-\!{\hat{\cal T}}^{mag}_J$
are expressible
as sums of products of single--particle matrix elements
(SPMEs) and one--body transition densities (OBTDs).
 
In calculating the nuclear matrix elements
we assumed harmonic oscillator wavefunctions
with an oscillator parameter $b = 1.94$~fm.
In addition, we assumed a constant
muon wavefunction over the
nuclear volume with 
$|\phi_{1s}|$$_{av}^2$ $=$ R$|\phi_{1s}(0)|^2$
where $\phi_{1s}(0)$ is the muon wavefunction
for a point-like nucleus and R$=$0.44 is the reduction factor 
for the $^{40}$Ca nucleus \cite{Wa75}. 
In calculating the muon capture rates
we generally fixed the numerically values of the weak vector, weak magnetic, weak axial 
and induced pseudoscalar couplings constants 
at $g_v$ $=$ 1.0, $g_m$ $=$ 3.706, $g_a$ $=$ 1.27
and $g_p$ $=$ 8.23. 
While the coupling $g_p$ is defined at 
q$^2$ $=$ $+$0.9m$_{\mu}^2$, the other couplings
are defined at q$^2$ $=$ 0
and therefore were scaled 
assuming a dipole form factor with $\Lambda^2$ $=$ 0.73 GeV$^2$. 
The momentum transfer $q$ was computed
via $q+q^2/2M_t$ $=$ $m_{\mu}$ $-$ $\Delta$E $-$ $\epsilon_b$
where 
M$_t$ is the target mass,
m$_{\mu}$ is the muon mass,
$\epsilon_b$ is the $\mu^-$ binding energy,
and $\Delta$E is the energy difference of the 
$^{40}$Ca-$^{40}$K nuclear states.

As mentioned above we employed harmonic oscillator 
wavefunctions in calculating nuclear matrix elements
-- such wavefunctions poorly reproducing 
the true asymptotic behavior  
of nuclear radial wavefunctions. 
We note the effect of different choices
of radial wavefunctions in first forbidden
beta-decay in $A \sim 40$ nuclei
was studied by Warburton {\it et al.}\ \cite{Wa88a}.
These authors reported only small differences  
between the relevant nuclear matrix elements obtained
with harmonic oscillator wavefunctions 
and Woods-Saxon wavefunctions.
In particular,
the d$_{3/2}$$\leftrightarrow$p$_{3/2}$ 
matrix elements relevant to
$(0^+ , 0)$$\rightarrow$$(0^- , 2626)$ capture
were found to differ by roughly 5-10\% 
(of course it is possible 
that their results are not applicable
at the higher momentum transfer 
of the muon capture process).
In addition, the sensitivity 
of weak matrix elements and muon capture rates
to different radial wavefunctions was studied
in $A = 12$-$32$ nuclei
by Kortelainen {\it et al} \cite{Ko00}.
Again, in most cases these authors observed only minor differences
between results derived 
with harmonic oscillator wavefunctions and 
Woods-Saxon wavefunctions. 

\section{Results for $( \mu^- , \nu )$ transition rates}
\label{s: rates}

In Table \ref{t: rates} we list the measured capture rates
and calculated capture rates for the relevant $( \mu , \nu )$
transitions to the low-lying $^{40}$K states.
We give the results of the WBMB and SAS calculations
and the Measday and Igo-Kemenes
experiments. 
Note that the experimental technique
of $\gamma$-ray detection
in Refs.\ \cite{Me06,Ig71} was insensitive
to the ground state transition
and the 30~keV state transition
and that Igo-Kemenes {\it et al.}\ 
only quoted results for the 
800, 1959 and 2047~keV states.
Also note that 
the 0~$\hbar$$\omega$ WBMB calculation implies
a vanishing rate to the positive-parity states
at 1632, 1959, 2260 and 2290~keV
and the SAS calculation gives no prediction
for the $d_{3/2}^{-1}$$p_{3/2}^{1}$ states
at 2047, 2070, 2104 and 2626~keV.

Overall -- for the ten measured rates
to the low-lying $^{40}$K states -- we claim
fair agreement between experimental data 
and model calculations from the perspective
to the global distribution of the capture rates, 
{\it i.e.} the presence
of strong, moderate and weak transitions.
Clearly though --
both in the comparison of experiment versus theory  
and model versus model
-- some discrepancies are obvious.
We give below the detailed assessment of the capture rates 
for the various allowed, first-forbidden 
and higher-forbidden transitions
In particular, we highlight the $(0^+ , 0)$$\rightarrow$$(0^- , 2626)$
axial charge transition,
the $(0^+ , 0)$$\rightarrow$$(2^- , 800)$ and
$(0^+ , 0)$$\rightarrow$$(2^- , 2047)$ unique first-forbidden
transitions, 
and the $(0^+ , 0)$$\rightarrow$$(1^+ , 2290)$
allowed Gamow-Teller transition.

\subsection{$0^+ \rightarrow 0^-$ first-forbidden transition}.

The capture rate for $0^+$$\leftrightarrow$$0^-$ transitions 
is determined by two nuclear matrix elements;
the axial charge matrix element $M_0 \sigma \cdot \nabla$,
which originates from the time component of the axial current,
and the $\ell = 1$ retarded Gamow Teller matrix element
$M_{01} \cdot \sigma$, which originates from the space component
of the axial current. 
The interest in $M_0 \sigma \cdot \nabla$
stems from its strong enhancement by pion exchange currents
while the interest in $M_{01} \cdot \sigma$ term
stems from the large contribution of the induced coupling $g_p$.
Recent studies of the $^{16}$O$(0^+, 0)$$\rightarrow$$^{16}$N$(0^-, 120)$
axial charge transition within  large-basis shell model calculations
have yielded values of $g_p = 6-9 $ \cite{Ha90} 
and $g_p = 7.5 \pm 0.5$ \cite{Wa94} that are
consistent with the theoretical prediction $g_p = 8.23$ 
from chiral symmetry arguments \cite{Go58,Be94,Fe97}.

\begin{figure}
\begin{center} 
\mbox{\epsfig{figure=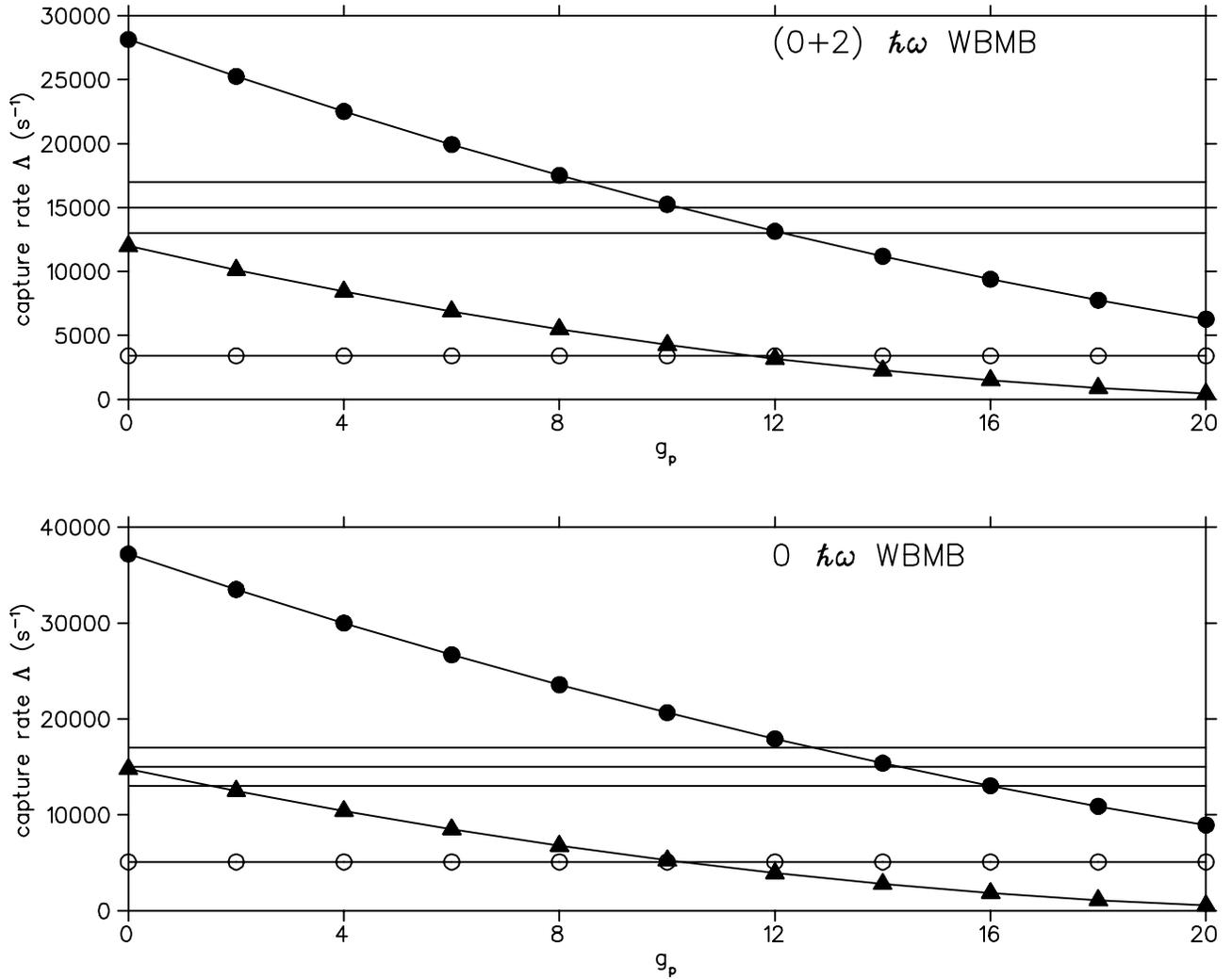,height=17.0cm,angle=90}}
\end{center}
\caption{Calculated rate for the $(0^+ , 0)$$\rightarrow$$(0^- , 2626)$ 
transition as a function of the coupling constant $g_p$
from the (0+2)~$\hbar$$\omega$ WBMB calculation (upper panel)
and the 0~$\hbar$$\omega$ WBMB calculation (lower panel).
The open circles correspond to the axial charge matrix element only,
the filled triangles correspond to the $\ell = 1$ retarded GT matrix 
element only, and the filled circles correspond to the full calculation.
The measured rate $\Lambda = (15 \pm 2) \times 10^3$~s$^{-1}$
is shown by the horizontal lines.}
\label{f: ac}
\end{figure}

Our calculated rate for the $(0^+ , 0)$$\rightarrow$$(0^- , 2626)$ 
transition as a function of the coupling $g_p$
is shown in Fig.\ \ref{f: ac}. The calculations were performed
with the 0~$\hbar$$\omega$ WBMB model 
and the (0+2)~$\hbar$$\omega$ WBMB model. 
Fig.\ \ref{f: ac} also shows the
results for (i) only the axial charge matrix 
element $M_0 \sigma \cdot \nabla$, (ii) only
the $\ell = 1$ retarded GT matrix element $M_{01} \cdot \sigma$, and (iii) 
both nuclear matrix elements. 
Note that the curves (i), (ii) and (iii) indicate the 
similar magnitude and constructive interference 
of the two contributions ({\it i.e.}\ $M_{01} \cdot \sigma$ and $M_{0} \sigma \cdot \nabla$)
to the $(0^+ , 0)$$\rightarrow$$(0^- , 2626)$ capture rate.
Using the experimental value of $\Lambda = ( 15 \pm 2 ) \times 10^3$~s$^{-1}$
for the $(0^+ , 0)$$\rightarrow$$(0^- , 2626)$  rate
the 0~$\hbar$$\omega$ WBMB calculation
yields a value of $g_p = 14.3^{+1.8}_{-1.6}$
and the (0+2)~$\hbar$$\omega$ WBMB calculation
yields a value of $g_p = 10.3^{+2.1}_{-1.9}$.
The former value is significantly larger than,
and the latter value is marginally consistent with, 
the theoretical prediction $g_p = 8.23$ \cite{Go58,Be94,Fe97}.

Note our calculations have 
omitted the effects of the exchange current renormalization
of the axial charge matrix element. 
Such effects are expected to enhance the 
matrix element $M_{0} \sigma \cdot \nabla$  
by typically a factor of two 
(see Ref.\ \cite{To86} for details).
A $\sim$50\% enhancement of the 
$M_0 \sigma \cdot \nabla$ matrix element
in the  $(0^+ , 0)$$\rightarrow$$(0^- , 2626)$ transition
would further increase the calculated capture rate 
and thereby further increase the induced coupling $g_p$. 

A crucial issue 
in the determination of the coupling $g_p$
from the $(0^+ , 0)$$\rightarrow$$(0^- , 2626)$ rate
is the model uncertainties in the $^{40}$Ca g.s.\ wavefunction.
As the multi-particle, multi-hole component
of the $^{40}$Ca ground state is increased
then the muon capture rate for the $(0^+ , 0)$$\rightarrow$$(0^- , 2626)$ 
transition is decreased, this explaining for the lower capture rate
and derived $g_p$-value 
in the (0+2)~$\hbar$$\omega$ WBMB calculation
and the higher capture rate 
and derived $g_p$-value 
in the 0~$\hbar$$\omega$ WBMB calculation.
As discussed later in Sec. \ref{ss: allowed},
we find that neither the 0~$\hbar$$\omega$ WBMB calculation
nor the (0+2)~$\hbar$$\omega$ WBMB calculation
give a satisfactory description
of the multi-particle, multi-hole admixtures
in the $^{40}$Ca ground state.
Unfortunately, 
we believe this result forestalls a 
firm conclusion on the in-medium value
of the induced pseudoscalar coupling $g_p$
from the $(0^+ , 0)$$\rightarrow$$(0^- , 2626)$ capture rate
(rather we suspect that the $^{40}$Ca g.s.\ wavefunction uncertainties 
are the most likely cause
of the unexpectedly large value 
of the coupling $g_p$ derived
from the $(0^+ , 0)$$\rightarrow$$(0^- , 2626)$ transition rate).

We note the effects of np-nh admixtures
in first forbidden beta-decay
were considered by Warburton {\it et al.}\ \cite{Wa88a}.
These authors concluded that the multi-particle multi-hole
components of the nuclear wavefunctions, while yielding 
relatively large corrections
for $\Delta$J$^{\pi}$ = 2$^-$ first forbidden transitions,
would yield relatively small corrections 
for $\Delta$J$^{\pi}$ = 0$^-$, 1$^-$ first forbidden transitions.
Specifically, they found an increase
of the $M_{0} \sigma \cdot \nabla$ axial charge matrix element 
and a decrease in the $M_{01} \cdot \sigma$ $\ell$-retarded 
GT matrix element of roughly 10\%,
a small correction that would largely cancel in the 
$(0^+ , 0)$$\rightarrow$$(0^- , 2626)$ capture rate. 
Herein, the apparent differences
between the 0~$\hbar$$\omega$ WBMB calculation
and (0+2)~$\hbar$$\omega$ WBMB calculation,
and different closed-shell occupancies
in the various model calculations,
is indicative of a significantly larger model uncertainty.

\subsection{$0^+ \rightarrow 2^-$ unique first-forbidden transitions}.

The experimental data also include two unique 
first forbidden $0^+ \rightarrow 2^-$ 
transitions to low-lying $^{40}$K states -- the (2$^-$, 800)
member of the $d_{3/2}^{-1}$-$f_{7/2}$
quadruplet and the (2$^-$, 2047)
member of the $d_{3/2}^{-1}$-$p_{3/2}$
quadruplet.
The model calculations indicate 
both transitions are governed by the weak axial coupling $g_a$
and the $\ell = 1$ matrix element $M_{21}$$\cdot$$\sigma$,
with the 
$( 0^+ , 0)$$\rightarrow$(2$^-$, 800) transition
being dominantly $d_{3/2}$$\rightarrow$$f_{7/2}$
in character 
and the $( 0^+ , 0)$$\rightarrow$(2$^-$, 2047) transition
being dominantly $d_{3/2}$$\rightarrow$$p_{3/2}$
in character. 

Concerning the $( 0^+ , 0)$$\rightarrow$(2$^-$, 800) transition
both model calculations and experimental data 
are in qualitative agreement on the very large capture rate. 
However, in detail the calculated rates are 
larger than the measured rates
by factors that range from $\sim$2 for 
the (0+2)~$\hbar$$\omega$ WBMB model
to $\sim$3 for the two other models.
Concerning the $( 0^+ , 0)$$\rightarrow$(2$^-$, 2047) transition
both model calculations and experimental data 
are in qualitative agreement on the rather moderate capture rate.
Unfortunately the experimental results  
are themselves only marginally consistent,
with the earlier result of Igo-Kemenes {\it et al.}\ 
being entirely consistent with the calculated rates
while the later result of Measday and Stocki being
somewhat larger than the calculated rates. 

For the first forbidden ($0^+$$,0)\rightarrow$(2$^-$, 800) transition
the analog transitions in 
electron scattering
$^{40}$Ca(e,$e^{\prime}$)$^{40}$Ca(8.43~MeV) \cite{Bu82,Og84,Pe95},
proton scattering 
$^{40}$Ca(p,$p^{\prime}$)$^{40}$Ca(8.43~MeV) \cite{Ej81}
and charge exchange
$^{40}$Ca(p,n)$^{40}$Sc(0.77~MeV) \cite{Ta84,Ch86}
are all well known.
In Table \ref{t: BM2} we compare the 
$0^+$$\rightarrow$$2^-$ reduced transition probability B(M2)
obtained from the $^{40}$Ca(e,$e^{\prime}$) data \cite{Pe95},
and the $0^+$$\rightarrow$$2^-$ capture rate $\Lambda$
obtained from the $^{40}$Ca$( \mu , \nu )$ data \cite{Me06}, 
with the corresponding predictions of our model calculations
and a pure $d_{3/2}$$\rightarrow$$f_{7/2}$ single-particle estimate.
The $d_{3/2}$$\rightarrow$$f_{7/2}$ single-particle estimates
are seen to exceed the
measured value of B(M2) by a factor 6.3 
and the measured value of $\Lambda$ by a factor 4.6.
By comparison the model calculations 
are seen to exceed the measured values 
of B(M2) by factors of 1.5-3.1
and the measured values of $\Lambda$ by factors of 1.6-2.8.
The similar quenching 
for the $0^+$$\rightarrow$$2^-$ transition
in the muon capture data and the electron scattering data
is not surprising, both processes being dominated 
by the matrix element of the spin-dipole ($M_{21}$$\cdot$$\sigma$) operator.
Comparable quenching is also reported  
for the analog transitions in the $^{40}$Ca(p,$p^{\prime}$) data
and the $^{40}$Ca(p,n) data.

This quenching or hindrance of the 
measured rate over the calculated rate --
as observed for the $( \mu^- , \nu )$ capture rate
and the $( e , e^{\prime} )$ reduced transition probability 
to the $(2^-, 800)$ state --
is well documented 
for unique first-forbidden $\beta$-decay 
in mass $A \sim 40$ nuclei (for example see 
Warburton {\it et al.}\ \cite{Wa88a}).
As documented in detail by Towner {\it et al.}\ \cite{To71}, 
such quenching originates from the repulsive character
of the $T = 1$ particle-hole interaction
and its coherent effects
on the particle-hole admixtures 
in the initial-final state wavefunctions.
In particular, in Ref.\ \cite{Wa88a} 
the authors report
a typical hindrance 
of approximately 3.7
for the known unique 
first-forbidden $\beta$-decay transitions 
in the $A \sim 40$ mass region
in a global comparison
to their WBMB calculations.

By comparison we observe no hindrance
of the measured capture rate 
compared to the calculated capture rate
for the ($0^+$,0)$\rightarrow$(2$^-$, 2047) transition. 
Interesting,
the ($0^+$,0)$\rightarrow$(2$^-$, 2047) transition
is dominantly $d_{3/2}$$\rightarrow$$p_{3/2}$
in character whereas the ($0^+$,0)$\rightarrow$(2$^-$, 800) transition
is dominantly $d_{3/2}$$\rightarrow$$f_{7/2}$
character (the unique first-forbidden beta decays
in the $A \sim 40$ mass region
are also $d_{3/2}$$\rightarrow$$f_{7/2}$ transitions).
As described by Towner  {\it et al.} \cite{To71}, 
the hindrances associated 
with $T =1$ particle-hole interactions are dependent on 
the orbital angular momenta of the particular transition,
and might differ between the $d$$\rightarrow$$f$ angular momenta 
of the ($0^+$,0)$\rightarrow$(2$^-$, 800) transition
and the $d$$\rightarrow$$p$ angular momenta 
of the ($0^+$,0)$\rightarrow$(2$^-$, 2047) transition
(of course it would be premature to draw 
any such conclusion
from a data-set
of two $0^+$$\rightarrow$$2^-$ transitions).

\subsection{$0^+ \rightarrow 1^+$ allowed Gamow-Teller transition}.
\label{ss: allowed}

The leading contribution to $0^+$$\rightarrow$$1^+$ transitions
originates from the weak axial coupling constant $g_a$ and
the spin-flip matrix element $M_{10}$$\cdot$$\sigma$.
Since the $M_{10}$$\cdot$$\sigma$ matrix element vanishes
in a simple closed-shell description
of the $^{40}$Ca ground state 
such $0^+$$\rightarrow$$1^+$ transitions
are especially sensitive 
to the np-nh admixtures
in the $^{40}$Ca g.s.\ wavefunction.

As discussed in Sec.\ \ref{s: 40Ca ground state} 
the specific np-nh admixtures in the $^{40}$Ca g.s.\
are strikingly different in the SAS model, 0~$\hbar$$\omega$ WBMB model,
and (0+2)~$\hbar$$\omega$ WBMB model.
In the SAS calculation
the $(0^+ , 0) \rightarrow ( 1^+ , 2290)$ transition
is dominated by a large $d_{3/2}$$\rightarrow$$d_{3/2}$ 
one-body transition density that originates
from the large $d_{3/2}^{-2}$-$f_{7/2}^{-2}$ admixture 
in the $^{40}$Ca g.s.\ wavefunction.
In the (0+2)~$\hbar$$\omega$ WBMB calculation
the $(0^+ , 0) \rightarrow ( 1^+ , 2290)$ transition
is comprised of many small one-body transition densities
that originate from
the fragmented np-nh admixtures
in the $^{40}$Ca g.s.\ wavefunction.
Consequently, 
the calculated $(0^+ , 0) \rightarrow ( 1^+ , 2290)$ 
rate is relatively large 
in the SAS model ($19.2 \times 10^3$~$s^{-1}$) 
and relatively small in the (0+2)~$\hbar$$\omega$ WBMB model
($0.08 \times 10^3$~$s^{-1}$).
The measured rate $(13 \pm 5) \times 10^3$~$s^{-1}$ is 
slightly smaller than the SAS prediction 
but much larger than the (0+2)~$\hbar$$\omega$ WBMB prediction,
thus suggesting a better treatment
of the np-nh admixtures
in the $^{40}$Ca g.s.\ wavefunction
by the SAS model.

This conclusion is supported
by the inelastic electron scattering data
for the $^{40}$Ca(e,$e^{\prime}$)$^{40}$Ca($1^+ , 9.87$~MeV) 
analog transition. This transition 
was studied by Petraitis {\it et al.}\ \cite{Pe95}
and yielded a reduced transition probability 
of B(M1)$ = 0.32 \pm 0.09$~$\mu_N^2$ 
to be compared with a SAS prediction of B(M1)$ = 0.86$~$\mu_N^2$ 
and a (0+2)~$\hbar$$\omega$ WBMB prediction of B(M1)$ \simeq 0.01$~$\mu_N^2$.
The moderate over-prediction of B(M1)
by the SAS model and 
gross under-prediction of B(M1)
by the WBMB model
is quite similar
to the corresponding comparisons
between the calculated rates
and the measured rates
in the $\mu$ capture reaction.
The similar scaling between 
experimental values and model values 
in $\mu$ capture and (e,$e^{\prime}$) scattering
is not surprising as both processes are 
dominated by the spin-flip $M_{10}$$\cdot$$\sigma$ 
matrix element.

As a final demonstration 
of the high sensitivity 
of the $(0^+ , 0)$$\rightarrow$$(1^+ , 0)$ transition
to the np-nh components
in the $A = 40$ wavefunctions
we list in Table \ref{t: ph} 
the $( 0^+ , 0)$$\rightarrow$$( 1^+, 2290)$ capture rate
versus the maximum occupancy of the $f_{7/2}$ orbital 
in the SAS model.
It clearly shows the strong correlation 
between the $0^+$$\rightarrow$$1^+$ transition rate
and the np-nh admixtures
in the $^{40}$Ca, $0^+$ ground state
and the $^{40}$K, $1^+$ excited state.
For example, by increasing the SAS model space
from 0-2~$\hbar$$\omega$ to 0-8~$\hbar$$\omega$
one increases the $0^+$$\rightarrow$$1^+$ capture rate 
from  $4.7 \times 10^3$~$s^{-1}$ to
$19.2 \times 10^3$~$s^{-1}$.

\subsection{Other transition}.

The remaining transitions comprise 
the  $(0^+ , 0)$$\rightarrow$$( 0^+ , 1632)$ 
non-analog Fermi transition, 
the   $(0^+ , 0)$$\rightarrow$$( 1^- , 2104)$ 
mixed first-forbidden transition,
and four $\ell > 1$ forbidden transitions
that range from second-forbidden to fifth-forbidden.

The leading contribution to $(0^+ , 0)$$\rightarrow$$( 0^+ , 1632)$ 
transition originates from the weak vector coupling constant 
$g_v$ and the nuclear matrix element $M_{0}$. 
As discussed earlier the transition rates for 
low-lying positive-parity states are highly sensitive 
to the multi-particle, multi-hole
admixtures in the $^{40}$Ca ground state.
In the SAS calculation the transition involves
the destructive interference of
$d_{3/2}$$\rightarrow$$d_{3/2}$ and
$f_{7/2}$$\rightarrow$$f_{7/2}$
s.p.\ transitions
and yields a capture rate $1.2 \times 10^3$~$s^{-1}$.
In the (0+2)~$\hbar$$\omega$ WBMB model the transition involves
many small contributions
from different s.p.\ transitions
that destructively interfere 
and yield a capture rate $0.01 \times 10^3$~$s^{-1}$.
Like the $( 0^+ , 0)$$\rightarrow$$( 1^+ , 2290 )$ transition
the large differences in calculated rates
are a consequence of the differences
in the  multi-particle, multi-hole admixtures 
in the $^{40}$Ca g.s.\ model wavefunctions.
The measured rate of
$(13 \pm 10 ) \times 10^3$~$s^{-1}$
has large uncertainties
and is consistent 
with both calculations.

The leading contribution to $( 0^+ , 0 )$$\rightarrow$$( 1^- , 2104 )$ 
transition originates from the weak axial coupling constant $g_a$ and
the nuclear matrix element $M_{11}$$\cdot$$\sigma$.
The transition to this $d_{3/2}^{-1}$-$p_{3/2}^{1}$ particle-hole
state is dominated by the $d_{3/2}$$\rightarrow$$p_{3/2}$
single particle transition. The measured rate of $(18 \pm 5) 
\times 10^3$~$s^{-1}$, 
0~$\hbar$$\omega$  WBMB rate of $14.7 \times 10^3$~$s^{-1}$, 
and (0+2)~$\hbar$$\omega$  WBMB rate of $9.0 \times 10^3$~$s^{-1}$,
are consistent within $1$-$2$ standard deviations.

The various model calculations 
for the four $\ell > 1$ forbidden transitions
all imply small capture rates.
While the measured rate for the fifth-forbidden $0^+$$\rightarrow$(5$^-$, 891)
transition and the third-forbidden $0^+$$\rightarrow$(3$^+$, 2260)
transition are reasonably consistent with model predictions,
the third forbidden $0^+$$\rightarrow$(3$^-$, 2070) transition
and second forbidden $0^+$$\rightarrow$(2$^+$, 1959) transition
have unexpectedly large measured rates 
for $\ell = 2, 3$ forbidden transitions.
We speculate -- as mentioned by Measday 
and Stocki \cite{Me06} --
that their experimental capture rates 
may include unidentified cascade feeding 
from higher-lying $^{40}$K levels.

\section{Recoil longitudinal alignment}
\label{s: alignments}

Both $0^+$$\rightarrow$$1^+$
and $0^+$$\rightarrow$$2^-$ transitions
involve contributions from the 
induced pseudoscalar coupling constant $g_p$.
While the contribution of $g_p$ 
to their capture rates is generally quite small
the contribution of $g_p$ 
to their longitudinal alignments 
are frequently rather large.\footnote{The 
recoil orientation about the neutrino momentum 
is termed the longitudinal orientation
and the recoil orientation about the muon spin
is termed the average orientation.
For $J \geq 1$ recoils the orientation
includes both a recoil polarization (rank-one 
orientation) and a recoil alignment 
(rank-two orientation). We consider the
longitudinal alignment $a_2$ which is 
experimentally accessible in the  $\gamma$-ray 
experiments.}
Given the observation of
$0^+$$\rightarrow$$1^+$ and $0^+$$\rightarrow$$2^-$ transitions
in $^{40}$Ca($\mu$,$\nu$) capture,
and determinations of
the analogous transition probabilities 
in $^{40}$Ca($e$, $e^{\prime}$) scattering,
we thought it interesting to consider the possibility
of determining $g_p$
by measuring the $^{40}$K alignment
following $^{40}$Ca capture.
Such measurements were previously performed
for muon capture on  $^{14}$N \cite{Go05},
$^{28}$Si \cite{Mo97} and $^{35}$Cl \cite{Ar02}.

\begin{figure}
\begin{center} 
\mbox{\epsfig{figure=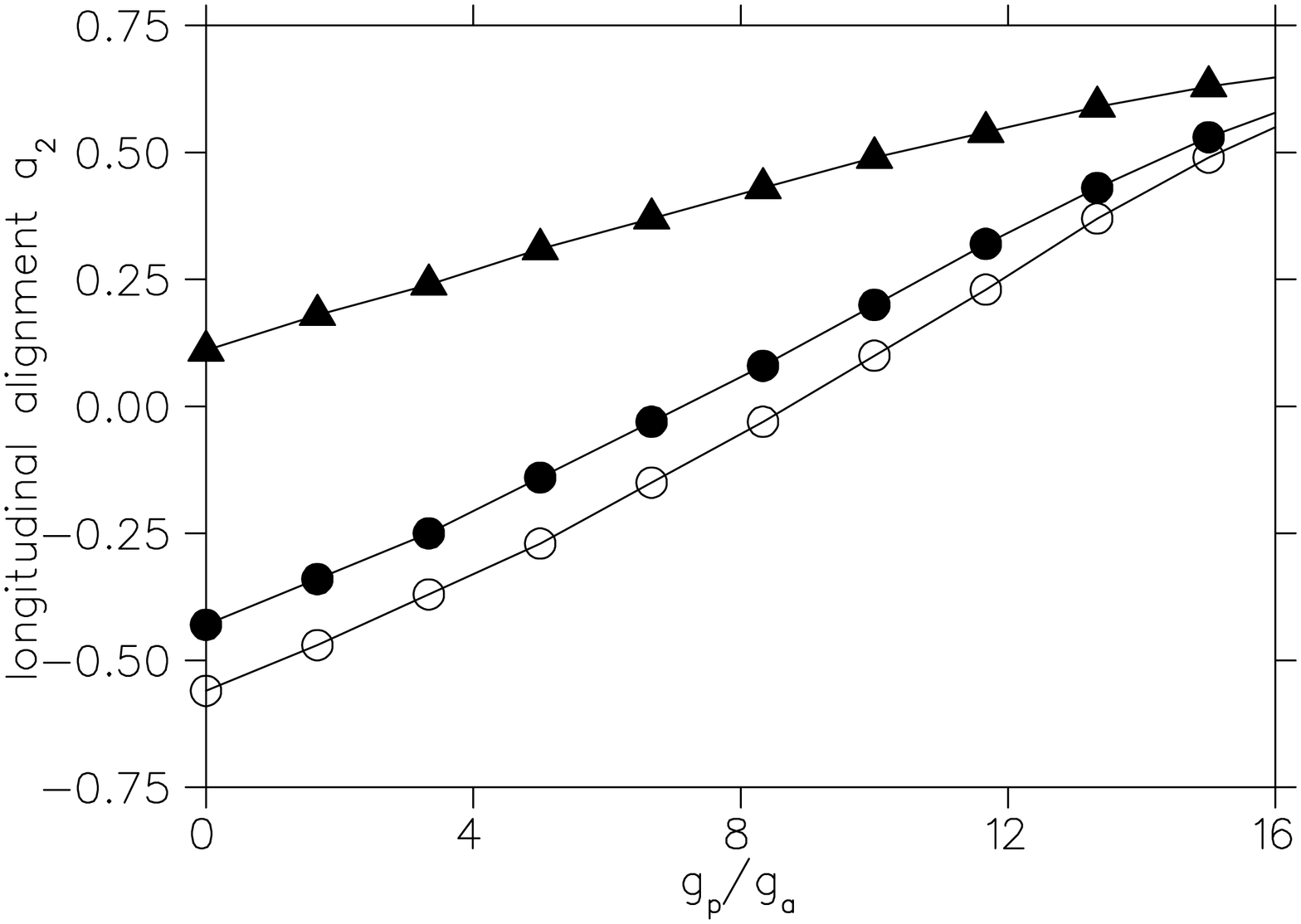,height=10.0cm,angle=90}}
\end{center}
\caption{Calculation of the longitudinal alignment $a_2$ for the 
$0^+$$\rightarrow$($1^+$,2290) transition versus
the coupling $g_p$. The solid circles indicate the SAS calculation
permitting up to eight particles in the $f_{7/2}$ orbital and 
the open circles indicate the SAS calculation
permitting up to four particles in the $f_{7/2}$ orbital.
The calculation employing only the $M_{10}$$\cdot$$\sigma$ matrix element
is denoted by triangles.}
\label{f: a2_1p}
\end{figure}

\begin{figure}
\begin{center} 
\mbox{\epsfig{figure=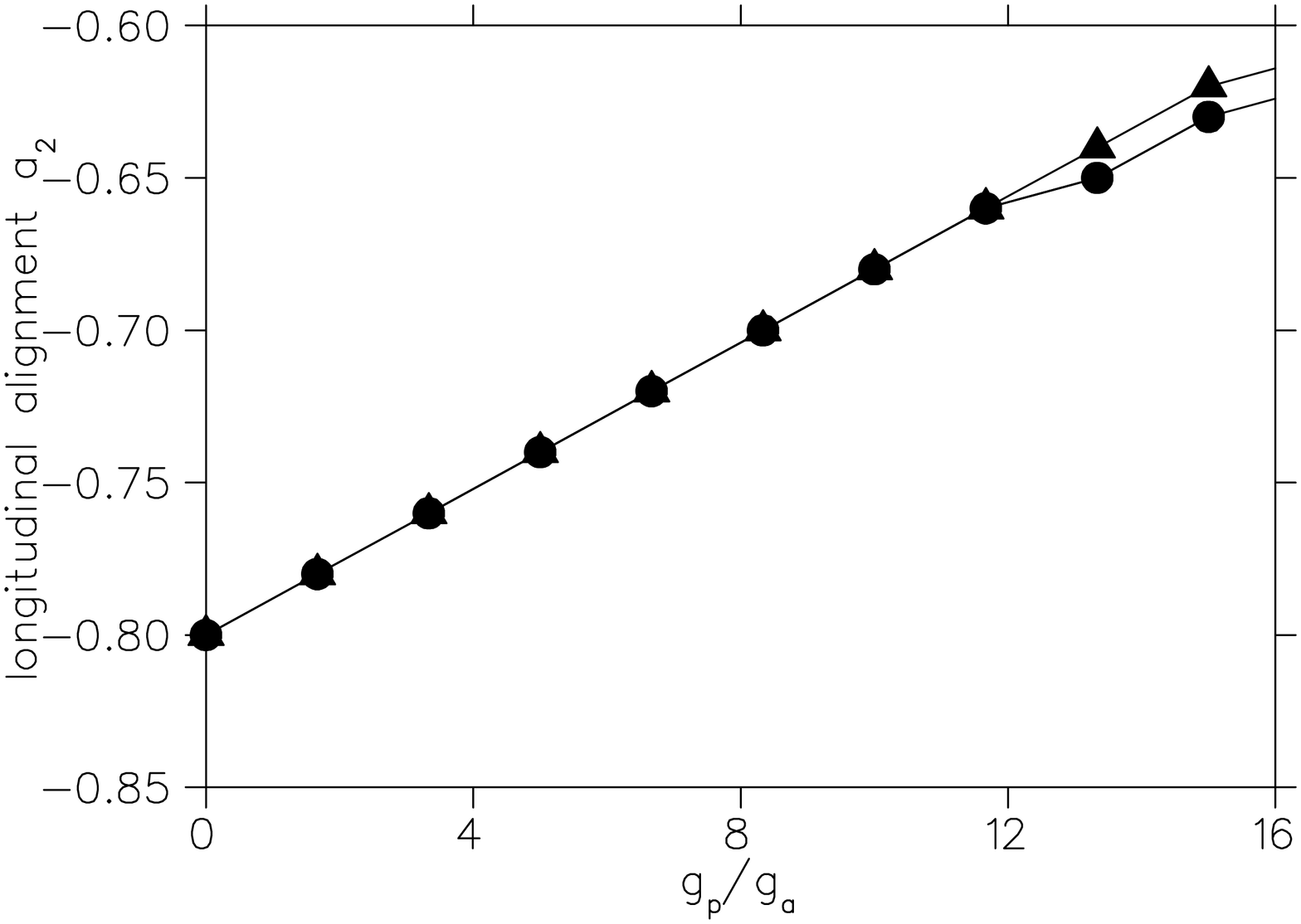,height=10.0cm,angle=90}}
\end{center}
\caption{Calculation of the longitudinal alignment $a_2$ for the 
$0^+$$\rightarrow$($2^-$,800) transition versus
the coupling $g_p$. The solid circles indicate the SAS calculation
permitting up to eight particles in the $f_{7/2}$ orbital and 
the open circles indicate the SAS calculation
permitting up to four particles in the $f_{7/2}$ orbital.
The calculation employing only the $M_{21}$$\cdot$$\sigma$ matrix element
is denoted by triangles.}
\label{f: a2_2n}
\end{figure}

Figs.\ \ref{f: a2_1p} and \ref{f: a2_2n} give results
for the longitudinal alignment $a_2$ 
versus the induced coupling $g_p$
for the $0^+ \rightarrow (2^- , 800)$ transition
and the $0^+ \rightarrow (1^+ , 2290)$ transition
using the SAS model (we adopt the SAS model rather than the
WBMB model as it reasonably reproduces both the capture rates).
Shown are results using 
(i) all the contributing nuclear matrix elements
and (ii) only the leading matrix elements
(either $M_{10}$$\cdot$$\sigma$ or $M_{21}$$\cdot$$\sigma$),
In addition, we plot the results with
maximum $f_{7/2}$ occupancies of either four nucleons 
or eight nucleons in order to gauge the model dependences.
Concerning the $0^+$$\rightarrow$$2^-$ transition
the calculation shows a relatively low sensitivity 
to both the coupling $g_p$ and the nuclear wavefunctions,
the wavefunction insensitivity reflecting the dominance 
of the $f_{7/2}$$^1$$d_{3/2}$$^{-1}$ transition 
and the $M_{21}$$\cdot$$\sigma$ matrix element.
Concerning the $0^+$$\rightarrow$$1^+$ transition 
the calculation shows
both greater sensitivity
to the coupling $g_p$
and greater sensitivity
to the nuclear wavefunctions,
the wavefunction sensitivity reflecting the
interference between $d_{3/2}$$\rightarrow$$d_{3/2}$
and $f_{7/2}$$\rightarrow$$f_{7/2}$ 
single-particle transitions and $M_{10}$$\cdot$$\sigma$ 
and $M_{1}$$\sigma$$\cdot$$\nabla$ matrix elements.

Could the longitudinal alignments
in $0^+ \rightarrow (2^- , 800)$ transitions
or $0^+ \rightarrow (1^+ , 2290)$ transitions
be measured?
In certain cases the alignment imparts
a directional correlation between the recoil nucleus 
and the subsequent de-excitation $\gamma$-ray \cite{Gr68},
and  is thereby measurable by the $\gamma$-ray Doppler lineshape.
However, such measurements require
both a suitable $\gamma$-decay spin sequence
and a short $\gamma$-decay lifetime.
Specifically, the $\gamma$-recoil
directional correlation $W( \theta )$ is given by 
\begin{equation}
W( \theta ) \propto 1 + a_2 B_{21} P_2(\cos{\theta})
\end{equation}
where $P_2(\cos{\theta})$ is the Legendre polynomial,
$\theta$ the angle between the recoil direction
and the $\gamma$-ray direction, and 
$B_{21}$ the $\gamma$-radiation  coefficient 
(see Ref.\ \cite{Ci84} for details).
For the $0^+$$\rightarrow$$1^+$ transition 
the dominant M1 decay ($1^+$,2290)$\rightarrow$($0^+$,1644) \cite{ENSDF}
yields $B_{21} = 1 / \sqrt{2}$ which implies a 
comparatively high sensitivity to $a_2$.
For the $0^+$$\rightarrow$$2^-$ transition 
the dominant M1 decay ($2^-$,800)$\rightarrow$($3^-$,30) \cite{ENSDF}
yields $B_{21} = 1 / \sqrt{70}$ which implies a 
comparatively low sensitivity to $a_2$.
In addition, the lifetime for the gamma de-excitation
must either be shorter or comparable
to the slowing-down time of the recoil ion.
Here the $0^+$$\rightarrow$$1^+$ lifetime of 83~fs
and the $0^+$$\rightarrow$$2^-$ lifetime of 280~fs 
are comparable to the recoil slowing-down time 
of $\sim$250~fs \cite{SRIM}. In short, we suspect
such measurements are possible but challenging.

\section{Summary}
\label{s: summary}

In summary we report the calculation
of partial rates and recoil orientations
in muon capture on calcium-40.
The calculations were performed with well-established,
well-tested nuclear models:
the WBMB interaction and
sd-pf model space of Warburton {\it et al.}\ \cite{Wa89}
and the SAS interaction and $d_{3/2}$-f$_{7/2}$ model space 
of Sakakura {\it et al.}\ \cite{Sa76}.
Moreover, the WBMB calculations were performed 
with both a simple closed-shell $^{40}$Ca
ground state and a mixed (0+2)~$\hbar$$\omega$ $^{40}$Ca ground state.
Taken together the calculations were capable of reproducing 
the important features of low-lying negative
and positive parity $^{40}$K levels.

Overall we observed fair agreement 
between measured capture rates and calculated capture rates 
for the low-lying $^{40}$K levels, {\it i.e.}\
the general distribution of muon capture rates to low-lying
$^{40}$K levels being reasonably consistent between the model
calculations and the experimental results.
We note however two striking exceptions -- 
the third forbidden $0^+$$\rightarrow$(3$^-$, 2070) transition
and second forbidden $0^+$$\rightarrow$(2$^+$, 1959) transition --
for which the measured rates exceeded
the calculated rates by very large factors.

Most importantly we emphasized the $(0^+ , 0)$$\rightarrow$$(0^- , 2626)$
axial charge transition and its high sensitivity 
to the induced pseudoscalar coupling $g_p$
of the proton's weak interaction.
Using the measured capture rate of Measday and Stocki \cite{Me06}
our (0+2)~$\hbar$$\omega$ WBMB calculation yielded 
$g_p = 10.3^{+2.1}_{-1.9}$ and our
0~$\hbar$$\omega$ WBMB calculation yielded
$g_p  = 14.3^{+1.8}_{-1.6}$.
Unfortunately, we concluded that neither 
the simpler 0~$\hbar$$\omega$ WBMB calculation
nor the  richer (0+2)~$\hbar$$\omega$ WBMB calculation
were capable of satisfactorily describing 
the multi-particle, multi-hole admixtures 
in the $^{40}$Ca ground state, 
thus forestalling a firm conclusion
on the in-medium value of the coupling constant $g_p$
from the $(0^+ , 0)$$\rightarrow$$(0^- , 2626)$ transition rate.

The above deficiencies 
in the model descriptions 
of the $^{40}$Ca ground state
were highlighted
by our discussion 
of the ($0^+$,0)$\rightarrow$$( 1^+, 2290 )$ allowed Gamow-Teller transition.
Such 0$^+$$\rightarrow$1$^+$ transitions
on doubly-magic $^{40}$Ca
are especially sensitive to
the np-nh admixtures
in the $^{40}$Ca ground state
as the $M_{10}$$\cdot$$\sigma$ spin-flip matrix element vanishes
for a simple closed-shell wavefunction.
Unfortunately, we found large model-to-model variations
in the different calculations
of the ($0^+$,0)$\rightarrow$(1$^+$, 2290) capture rate,
and most worrisome for the analysis
of the $(0^+ , 0)$$\rightarrow$$(0^- , 2626)$ transition,
the (0+2)~$\hbar$$\omega$ WBMB calculation grossly under-estimates
the ($0^+$,0)$\rightarrow$(1$^+$, 2290) transition rate.

In addition, we discussed two
$0^+$$\rightarrow$$2^-$ unique first forbidden transitions
to $^{40}$K levels at 800 and 2047~keV.
In nuclear beta-decay such first-forbidden transitions
have been extensively studied in the 
context of their hindrance
via the coherent effects
of the repulsive $T$=$1$ 
particle-hole interaction.
Intriguingly, we 
found a substantial hindrance of the $(0^+ , 0)$$\rightarrow$$(2^- , 800)$
transition that has dominant $d_{3/2}$$\rightarrow$$f_{7/2}$ character
but a negligible hindrance of the $(0^+ , 0)$$\rightarrow$$(2^- , 2047)$
transition that has dominant $d_{3/2}$$\rightarrow$$p_{3/2}$ character. 
The latter case of a $d_{3/2}$$\rightarrow$$p_{3/2}$ transition 
is generally inaccessible via the $\beta$-decay studies
in the $A \sim 40$ mass region.

A number of improvements could be made on our work.
The calculations were conducted 
using limited model spaces, harmonic
oscillator nuclear wavefunctions, and uniform muonic wavefunctions.
In particular, 
the multi-particle, multi-hole structure 
of $^{40}$Ca ground state
was a major difficulty
for the model calculations,
it limiting 
our conclusions
for the induced coupling $g_p$
in the $(0^+ , 0)$$\rightarrow$$(0^- , 2626)$ transition.
Given the sensitivity
of the  $0^+$$\rightarrow$$0^-$ axial charge transition
to the induced pseudoscalar coupling,
the interest 
in the  $0^+$$\rightarrow$$2^-$ unique first forbidden transitions
due to core polarization effects,
and the interest 
in the  $0^+$$\rightarrow$$1^+$ allowed Gamow-Teller transitions
due to ground state correlations,
we strongly encourage further theoretical work
on $^{40}$Ca muon capture.
In addition,
new experimental efforts
on exclusive muon capture
in other $A \sim 40$ nuclei
would help to extend 
the experimental data-set
and benefit any model studies.

We wish to thank both Prof.\ David Measday and Dr.\ Trevor Stocki
for valuable discussions 
and gratefully acknowledge the National Science
Foundation (USA) for their financial support.

\newpage

\begin{table*}
\caption{Comparison of measured rates and calculated rates
for muon capture to low-lying $^{40}$K levels. 
The first two columns list the spin-parities and excitation energies
of the $^{40}$K states and the final column lists the dominant 
particle-hole configurations of the $^{40}$K states.
The experimental rates are taken from the experimental work
of Measday and Stocki (column three) and Igo-Kemenes {\it et al} 
(column four) and the calculated rates were obtained with the
SAS, 0~$\hbar$$\omega$ WBMB, and (0+2)~$\hbar$$\omega$ WBMB models 
(see text for details).}
\label{t: rates}
\begin{center}
\begin{tabular}{llcccccc}
  & & & & & & & \\ 
 $J_f^{\pi}$  & 
E$_x$ & Exp.\ \cite{Me06} & Exp.\ \cite{Ig71} &   
 SAS rate     &
 0~$\hbar$$\omega$ WBMB   & (0+2)~$\hbar$$\omega$ WBMB  & dominant \\
 & 
(keV) & ($10^3$~s$^{-1}$)  &  ($10^3$~s$^{-1}$)  &   
 ($10^3$~s$^{-1}$)    &
 ($10^3$~s$^{-1}$) & ($10^3$~s$^{-1}$) & p-h config.\   \\
  & & & & & & & \\ 
\hline
  & & & & & & & \\ 
 4$^-$ &    0 &              &            &  3.2  &   2.7 &  1.7 & $(d_{3/2})^{-1}$$(f_{7/2})^{1}$ \\
 3$^-$ &   30 &              &            &  7.0  &   3.4 &  2.1 & $(d_{3/2})^{-1}$$(f_{7/2})^{1}$ \\
 2$^-$ &  800 &  127$\pm$13  & 108$\pm$30 &  339  &   351 &  200 & $(d_{3/2})^{-1}$$(f_{7/2})^{1}$ \\
 5$^-$ &  891 &  5.1$\pm$2.5 &            &  0.0  &   0.0 &  0.0 & $(d_{3/2})^{-1}$$(f_{7/2})^{1}$ \\
 0$^+$ & 1632 &  13$\pm$10   &            &  1.2  &   0.0 &  0.0 & $(d_{3/2})^{-2}$$(f_{7/2})^{2}$ \\
 2$^+$ & 1959 &  31$\pm$5    &   13$\pm$5 &  5.8  &   0.0 &  0.2 & $(d_{3/2})^{-2}$$(f_{7/2})^{2}$  \\
 2$^-$ & 2047 &  21$\pm$7    &   12$\pm$4 &       &  11.5 &  7.4 & $(d_{3/2})^{-1}$$(p_{3/2})^{1}$ \\  
 3$^-$ & 2070 &  18$\pm$7    &            &       &   0.8 &  0.6 & $(d_{3/2})^{-1}$$(p_{3/2})^{1}$ \\
 1$^-$ & 2104 &  18$\pm$5    &            &       &  14.7 &  9.0 & $(d_{3/2})^{-1}$$(p_{3/2})^{1}$ \\
 3$^+$ & 2260 &  $<$6        &            &  0.0  &   0.0 &  0.0 & $(d_{3/2})^{-2}$$(f_{7/2})^{2}$  \\
 1$^+$ & 2290 &  13$\pm$5    &            & 19.2  &   0.0 &  0.08 & $(d_{3/2})^{-2}$$(f_{7/2})^{2}$  \\
 0$^-$ & 2626 &  15$\pm$2    &            &       &  23.2 & 17.2 & $(d_{3/2})^{-1}$$(p_{3/2})^{1}$ \\ 
  & & & & & & & \\ 
\end{tabular}
\end{center}
\end{table*}

\newpage

\begin{table*}
\caption{Comparison of measured and calculated muon capture rates $\Lambda$,
and measured and calculated reduced transition probabilities B(M2),
for the ($0^+$,0)$\rightarrow$$( 2^- ,800 )$ unique first forbidden transition.
Column three lists the $d_{3/2}$$\rightarrow$$f_{7/2}$ single-particle estimate
and columns four, five and six list
the results of the SAS, 0~$\hbar$$\omega$ WBMB and
(0+2)~$\hbar$$\omega$ WBMB models (see text for details).}
\label{t: BM2}
\begin{center}
\begin{tabular}{cccccc}
 & & & & & \\ 
observable  & measured &
$d_{3/2}$$\rightarrow$$f_{7/2}$   & 
SAS  & 0~$\hbar$$\omega$ WBMB & (0+2)~$\hbar$$\omega$ WBMB \\   
 & value & s.p.\ estimate & model & model & model \\
 & & & & & \\ 
\hline
 & & & & & \\
B(M2) ($\mu_N$$^2$ fm$^2$) &  235$\pm$20 & 1495 & 778 & 692 & 375 \\
$\Lambda$ ($\times$10$^3$~s$^{-1}$) & 127$\pm$13 & 600 & 339 & 351 & 200 \\
 & & & & & \\ 
\end{tabular}
\end{center}
\end{table*}

\begin{table*}
\caption{The calculated muon capture rate for the
($0^+$,0)$\rightarrow$(1$^+$,2290~keV) allowed Gamow-Teller transition
versus the maximum occupancy of the $f_{7/2}$ occupancy.
Also given are the 2p-2h and 4p-4h admixtures 
in the $^{40}$Ca and $^{40}$K wavefunctions.}
\label{t: ph}
\begin{center}
\begin{tabular}{cccccc}
 & & & & & \\ 
maximum  & $^{40}$Ca 2p-2h  & $^{40}$Ca 4p-4h (\%) 
& $^{40}$K 2p-2h & $^{40}$K 4p-4h & rate \\
 $f_{7/2}$ occupancy & admixture (\%) 
& admixture (\%) & admixture (\%) & admixture (\%) 
& ($\times$10$^3$~s$^{-1}$) \\ 
 & & & & & \\ 
\hline
 & & & & & \\
 2 & 27.0 & & 100 & & 4.7 \\
 4 & 35.7 & 8.9 & 71.3 & 28.7 & 11.0 \\
 6 & 35.5 & 13.6 & 56.5 & 36.0 & 16.8 \\
 8 & 35.0 & 14.5 & 52.5 & 36.3 & 19.2 \\
 & & & \\ 
\end{tabular}
\end{center}
\end{table*}

\newpage



\newpage

\newpage

\newpage

\end{document}